\documentclass[final,5p,times,twocolumn]{elsarticle}

\usepackage{amssymb}

\journal{Journal of Crystal Growth}

\begin{document}

\begin{frontmatter}



\title{High-pressure melt growth and transport properties of SiP, SiAs, GeP, and GeAs 2D\,layered semiconductors}


\author{C. Barreteau\fnref{1}}
\author{B. Michon}
\author{C. Besnard}
\author{E. Giannini}
\address{Department of Quantum Matter Physics,
University of Geneva, 24 Quai Ernest-Ansermet, 1211 Geneva,
Switzerland}

\fntext[1]{Celine.Barreteau@unige.ch}

\begin{abstract}
Silicon and Germanium monopnictides SiP, SiAs, GeP and GeAs form a
family of 2D layered semiconductors. We have succeeded in growing
bulk single crystals of these compounds by melt-growth under high
pressure (0.5-1\,GPa) in a cubic anvil hot press. Large (mm-size),
shiny, micaceous crystals of GeP, GeAs and SiAs were obtained, and
could be exfoliated into 2D flakes. Small and brittle crystals of
SiP were yielded by this method. High-pressure sintered
polycrystalline SiP and GeAs have also been successfully used as a
precursor in the Chemical Vapor Transport growth of these crystals
in the presence of I$_{2}$ as a transport agent. All compounds are
found to crystallize in the expected layered structure and do not
undergo any structural transition at low temperature, as shown by
Raman spectroscopy down to T=5\,K. All materials exhibit a
semiconducting behavior. The electrical resistivity of GeP, GeAs
and SiAs is found to depend on temperature following a 2D-Variable
Range Hopping conduction mechanism. The availability of bulk
crystals of these compounds opens new perspectives in the field of
2D semiconducting materials for device applications.

\end{abstract}

\begin{keyword}
A2. High-pressure melt growth \sep B2. Semiconducting materials
\sep A1. Low dimensional structures \sep A2. Growth from vapor




\end{keyword}

\end{frontmatter}


\section{Introduction}
\label{1}

2D materials are of great interest for the novel electronic
properties that can arise from the reduced dimensionality and the
quantum confinement of charge carriers, and have become more and
more appealing for applications in modern electronic devices.
After the epoch-making discovery of graphene \cite{Novoselov04},
the search for stable free-standing atomic layers of
semiconducting materials has experienced a rush and a fast
improvement of the processing techniques. The wide family of
transition metal dichalcogenides (TMDs) has proven to be the most
promising, offering quite a large variety of compounds, large
tunability of properties and flexibility in potential practical
applications \cite{Chhowalla13}. Electronic and optoelectronic
devices based on various TMDs have been demonstrated
\cite{Radisavljevic11,Wang12,Tosun14}.
The search for other families of 2D materials exhibiting the same
properties, existing in stable atomic layers and offering similar
potential for applications together with a natural abundance and a
low production cost is still very active and deserves a strong
effort. Chemically stable atomic layers with no surface dangling
bonds can be obtained from other layered materials and the van der
Waals-like bond between layers with different chemical
compositions opens new perspectives for new heterostructures to be
realized in a wide range of materials. Besides graphene, examples
of pure elements from group IV (Si and Ge) and group V (P) have
been found to form atomically thin layers (silicene \cite{Vogt12},
germanene \cite{Davila14} and phosphorene \cite{Li14},
respectively) that can be obtained through either chemical
deposition on substrate \cite{Lalmi10} or mechanical exfoliation
of bulk 3D crystals \cite{Kara12}. Binary compounds of a group IV
element (Si, Ge, Sn) and a group V pnictogens (P, As,…) are also
known to form layered structures in which 2D strongly covalent
layers are stacked onto each other through weak van der Waals like
bonds, as well as in TMDs. Silicon and germanium phosphides and
arsenides have been reported since decades to crystallize in
various layered structures with either orthorhombic
(\emph{C}\emph{mc}2$_{1}$ space group, SiP \cite{Wadsten75},
\emph{Pbam} space group, SiP$_{2}$ and GeAs$_{2}$
\cite{Wadsten67}) or monoclinic (\emph{C}2/\emph{m} space group,
GeP, GeAs and SiAs \cite{Wadsten67,Mentzen81,Wadsten65})
symmetries. After the initial investigations of their crystal
structures and phase equilibria, during the sixties and seventies,
this family of compounds has been rather overlooked, and attracts
today our interest as being a potential class of 2D materials,
alternative to TMDs. The equilibrium phase diagrams assessed so
far predict the existence of a limited number of stable
compositions and polytypes \cite{Villars02}. Few more have been
synthesised at high pressure (cubic GeP, cubic GeP$_{2}$,
rhombohedral GeP$_{3}$ \cite{Osugi67,Gullman72}) or suggested to
exist according to structural investigations, even though not
present in the equilibrium phase diagram (orthorhombic
SiP$_{2}$\cite{Wadsten67} and cubic SiP$_{2}$ \cite{Wadsten67b}).
Recently, the phase diagram of the Si-P has been theoretically
revisited under high pressure and suggested to be substantially
different from that drawn under equilibrium conditions at ambient
pressure \cite{Liang14}. First principle calculations of phase
stability in the Si-P system have predicted the existence of at
least three new stable Si$_{x}$P$_{y}$ compounds with a layered
structure that could be stable in single atomic layer forms
\cite{Huang15}. Bulk crystals of these materials have been seldom
if ever grown: SiAs was reported to crystallize from the melt by
Sudo \cite{Sudo80} and from the vapor phase by Kutzner et al.
\cite{Kutzner11}; SiP and SiAs crystals were grown by the physical
vapor transport (PVT) method \cite{Beck66}, but resulted not to
crystallize in the expected space group; very recently GeP was
reported to grow in crystalline form by using a solution growth
method in a flux of Bi and Sn \cite{Lee15}. As a matter of fact,
the volatility and the strong reactivity and toxicity of
pnictogens require the use of close reactors in order to prevent
the vapor phase from escaping. Here we report about crystal growth
of the four members of the family, namely SiP, SiAs, GeP and GeAs,
from the self-flux under high pressure, using a cubic anvil hot
pressure apparatus. Large, micaceous, and easy-to-cleave crystals
were obtained in the case of GeP, GeAs, and SiAs. Small and
brittle crystals were obtained in the case of SiP. Polycrystalline
binary samples, processed in the same high-pressure furnace, were
used as a precursor material for CVT growth experiments with
iodine as a transport agent. This method was found to favor the
growth of SiP and GeAs. This article reports on materials
processing, crystal growth, structural and physical
characterization of SiP, SiAs, GeP and GeAs. The crystals have the
expected layered structure and can be exfoliated. These materials
exhibit semiconducting behavior and confirm to have high potential
as 2D materials for novel nano-engineered semiconducting devices.

\section{Experimental}
\label{2}

\subsection{Thermodynamic considerations}
\label{2.1}

The phase diagrams of the four systems under investigation have
been assessed and are reported in the Pauling Files database
\cite{Villars02}. Only two stable compounds are reported to exist
in the Si-P and Ge-P diagrams, SiP and GeP, whose decomposition
occurs at ~1160$^\circ$C and 750$^\circ$C, respectively, via
peritectic decomposition into elements P and Si, or Ge. According
to those diagrams, SiP is reported to transform into a mixed
solid-liquid-vapor phase \cite{Giessen59}, whereas GeP is claimed
to decompose into solid Ge and liquid P \cite{Ugai78}. At ambient
pressure this appears quite unlikely, the sublimation temperature
of P being as low as 430$^\circ$C. The Si-P phase diagram has been
recently refuted and corrected by Liang et al. \cite{Liang14}. On
the other hand, no recent thermodynamical investigations of the
Ge-P system, or diagram updates, have been undertaken.
Undoubtedly, the decomposition into a vapor phase of phosphorous
at ambient pressure prevents from processing SiP and GeP by
conventional techniques.

The phase diagrams of the systems containing As have been
investigated by various authors (see \cite{Villars02} for a
complete collection), and all agree on a congruent melting of SiAs
and GeAs into a liquid phase at ambient pressure. This allows
processing the compounds and growing the crystals under more
conventional conditions. As a matter of fact, arsenic melts into
liquid at ambient pressure and its vapor pressure is more than two
orders of magnitude lower than that of P at the same temperature.
In the case of SiAs and GeAs, the growth from the melt can be made
more difficult by the presence of monoarsenide and diarsenide
phases, both congruently melting in the same temperature range,
which can grow into one another in the case of composition
fluctuations and element segregation. Crystals of SiAs have been
grown successfully from the melt under ambient pressure
\cite{Beck66}, as well as from the vapor phase under vacuum
\cite{Kutzner11}. Only $\mu$m-size crystals of GeAs have been
obtained so far \cite{Mentzen81}.

As a result of the above considerations, we have chosen a
melt-growth method under high-pressure (in the GPa range) for all
compounds. The calculated phase diagram of Si-P at 0.1 GPa
\cite{Liang14}, showing a solid-liquid equilibrium at the SiP
composition, supports this choice. Moreover, high pressure has
also been successfully used for growing crystals of black
phosphorous preventing both its sublimation and the transformation
into dangerous white phosphorous \cite{Endo82}. The use of
pressures as high as 0.5-1 GPa is expected to make the growth of
SiP and GeP from the melt possible. In the presence of As, the
high pressure impedes the toxic vapor of As from reacting with the
atmosphere, and speeds up the growth kinetics.

\subsection{High-pressure melt growth (HP)}
\label{2.2}

All crystals were grown in a high pressure cubic anvil press. Pure
elements Si (6N), Ge (5N), P (5N) and As (5N) were used as
reactants. They were mixed in a stoichiometric ratio and pressed
into pellets of approximately 7\,mm of diameter and 3\,mm of
thickness under uniaxial stress (3\,tons). The pellets were then
placed in a cylindrical boron nitride crucible surrounded by a
graphite sleeve resistance heater and inserted into a pyrophyllite
cube as a pressure transmitting medium. The pyrophyllite cell was
then placed inside the high-pressure set-up, which consists of six
WC anvils. For each composition, the cell was cold pressurized,
then fast brought to high temperature (T$_{1}$) at
1200$^\circ$C/hours, held for 30\,min at this temperature and then
slowly cooled to a temperature T$_{2}$ before being quenched to
room temperature, while maintaining a constant pressure (see Table
1 for details). The temperature T$_{1}$ was chosen to be above the
complete melting of the precursors. The quench temperature was
chosen to be above any possible decomposition or phase transition.
The slow cooling rate allows crystals to nucleate and grow.

\begin{table}[h!]
  \begin{tabular}{|c|c|c|c|c|}
    \hline
 Sample&Pressure&T$_{1}$($^\circ$C)&T$_{2}$($^\circ$C)&Cooling ($^\circ$C/h)\\
    \hline\hline
 SiP&0.5/1 GPa&1050&925&25\\
 SiAs&1 GPa&1200&1000&50\\
 GeP&1 GPa&900&600&100\\
 GeAs&1 GPa&900&650&50\\
     \hline
\end{tabular}
\caption{HP growth conditions}

\end{table}

\subsection{Chemical Vapor Transport (CVT)}
\label{2.3}

As mentioned above, vapor transport techniques are the most common
way to obtain single crystals of these materials. However, owing
to the large difference between the vapor pressure of the
pnictogen and the group-IV element, very large temperature
gradients ($\sim$25$^\circ$C/cm) \cite{Wadsten69} were employed to
achieve the right stoichiometry. In order to reduce such a strong
technical constraint, we tried to grow crystals by vapor transport
using smaller temperature gradients (5-7$^\circ$C/cm).

For the Physical Vapor Transport (PVT), the pure elements were
mixed in a stoichiometric ratio, with a total mass of 0.2-0.3\,g.
The mixture was then placed in a quartz ampoule with an internal
diameter of 8\,mm and a length of 120\,mm and sealed under vacuum
(5x10$^{-6}$ mbar). For the Chemical Vapor Transport (CVT), a
transport agent (I$_{2}$) was added to the pure elements according
to a molar ratio n(I$_{2}$)/n(Group IV)\,=\,0.05. In both cases,
the sealed reactor was placed in a two-zone furnace in the
presence of a thermal gradient dT/dx $\approx$ 5-7$^\circ$C/cm,
and heated up to T$_{hot}$ at the hot end, equal to 1100$^\circ$C
and 900$^\circ$C for SiP and GeAs, respectively. After few days,
the furnace was switched off and the temperature decreased to room
temperature. Preliminary results confirmed the difficulties in
maintaining the wanted stoichiometry during crystallization in
both techniques, more dramatically in the PVT case, due to the
rapid sublimation of the pnictogen element during heating. The
addition of the transport agent proved to be insufficient to
successfully control the growth process. According to these
observations, we decided to start from different reactants and
grow single crystal only by the CVT technique. Instead of using a
mixture of pure elements, we started from high-pressure
pre-reacted binary precursors mixed with the transport agent. This
processing route proved to be successful to grow single crystals
of SiP and GeAs.

\begin{figure}[b!]
\includegraphics[width=\columnwidth]{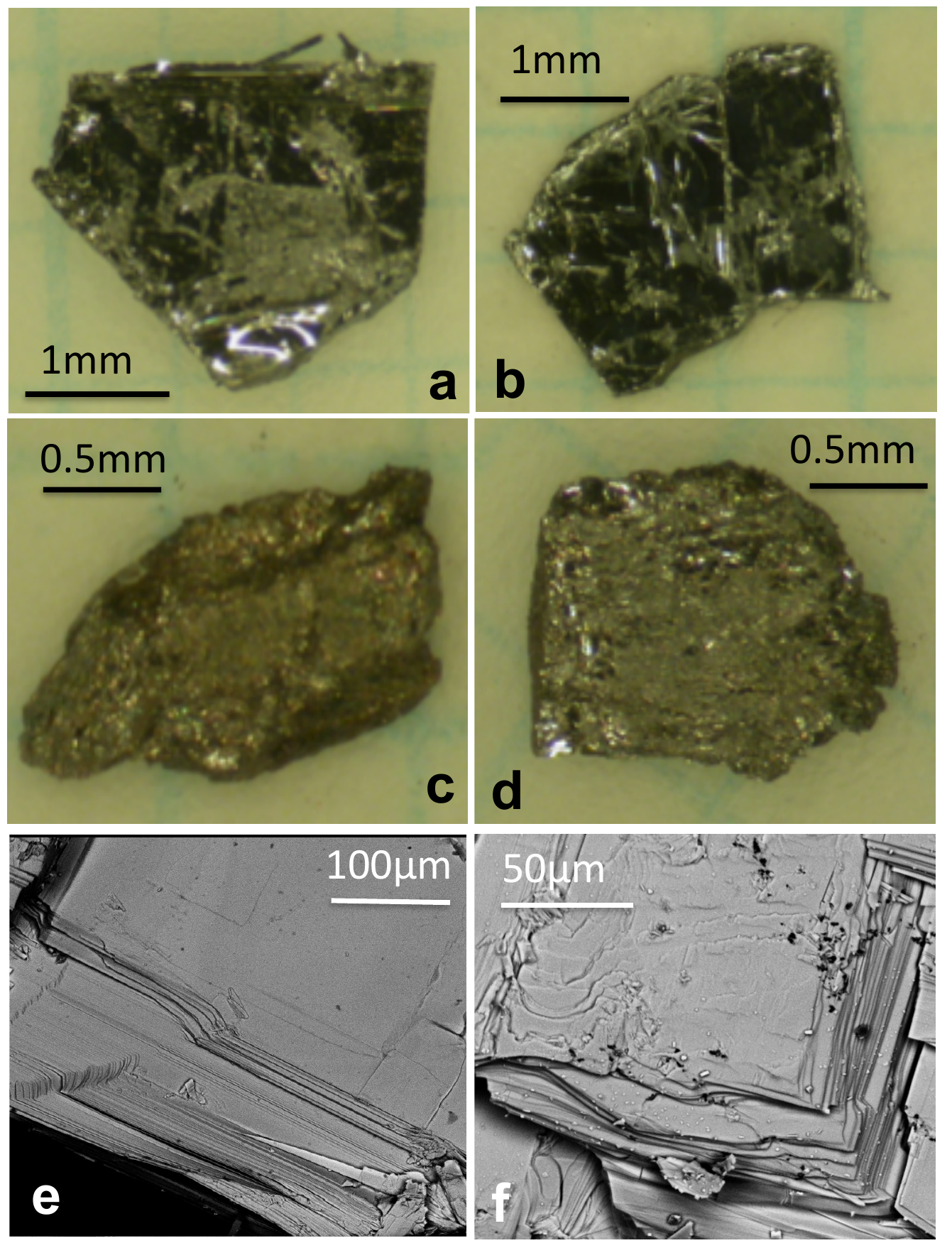}
\caption{Crystals of GeP (a), GeAs (b), SiP (c) and SiAs (d) from
HP melt growth; SEM images of GeAs (e) and SiAs (f) evidencing the
lamellar structure and the easy cleavability.
\label{fig:BarreteauJCG_Fig1}}
\end{figure}

\subsection{Structural and Physical characterization}
\label{2.3}

As-grown crystals were characterized by X-Ray diffraction (XRD),
SEM-EDX analysis, Raman Spectroscopy, and electrical transport.
XRD patterns were acquired in a Philips X'Pert four-circle
diffractometer and a Philips PW1820 powder diffractometer, both
using Cu K$\alpha$ radiation. Thanks to their better
quality, only GeP and GeAs crystals could also be measured in an
Agilent Super Nova single crystal diffractometer using MoK$\alpha$
radiation. SEM-EDX analysis were carried out in a LEO 438VTP
electron microscope coupled to a Noran Pioneer X-Ray detector.
Raman spectroscopy was performed with a homemade micro Raman
spectrometer equipped with an argon laser ($\lambda$ = 514.5\,nm,
spectral resolution of $\approx$1\,cm$^{-1}$) and a helium-flow
cryostat  working from 300K to 4\,K. The electrical resistivity
was measured by the standard four-probe method using a Quantum
Design PPMS (physical properties measurement system) from 300\,K
to 2\,K under a magnetic field from 0 to 7\,T.

\section{Results and Discussion}
\label{3}

Pictures of single crystals of SiP, SiAs, GeP and GeAs grown by
the HP method are presented in Fig.\,\ref{fig:BarreteauJCG_Fig1}.
As shown in these pictures, it is easier to grow the pnictides of
germanium than those of silicon. Single crystals of GeP, GeAs and
in a lesser measure SiAs are large, shiny and present grey flakes
that can be easily cleaved into thinner flakes. On the other hand,
SiP crystals are significantly smaller, less shiny, and very
brittle. The peculiar lower quality of SiP crystals is ascribable
to various facts: SiP is the only monopnictide that crystallizes
in the orthorhombic space group $\emph{Cmc}$2$_{1}$ instead of the
monoclinic $\emph{C}$2/$\emph{m}$, which is common to the other
compositions. Moreover, according to the phase diagram reported by
Liang et al. \cite{Liang14}, the temperature range suitable for
nucleation and growth of SiP is rather narrow. Besides, the
isostatic pressure can only be increased over a little range for
avoiding the formation of other HP metastable Si$_{x}$P$_{y}$
phases \cite{Osugi66}. Congruent melt conditions were achieved for
all systems: SEM-EDX analysis confirms the expected 1:1 chemical
composition of the crystals and neither composition fluctuations
nor secondary phases have been noticed in the core of the HP-grown
bulk. SEM images, figure 1(e-f), evidence the lamellar structure
of these crystals and their suitability for fabricating 2D
devices. The same analysis on the GeAs and SiP crystals grown by
the CVT method also confirms the 1:1 stoichiometry, with no traces
of the transport agent I$_{2}$.

\begin{figure}[t!]
\includegraphics[width=\columnwidth]{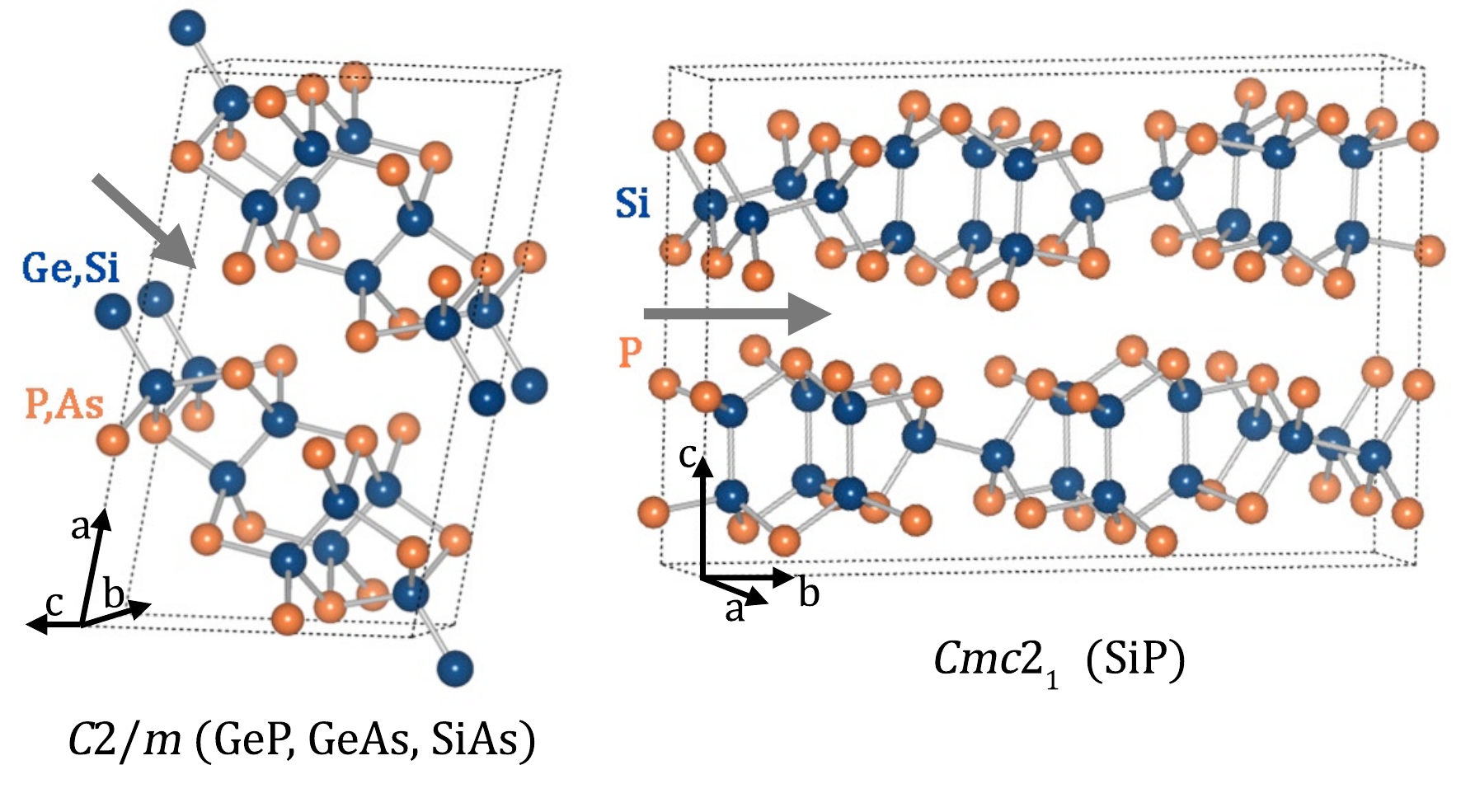}
\caption{Crystal structure of monoclinic GeP, GeAs and SiAs (left)
and orthorhombic SiP (right). Grey arrows indicate the cleavage
plane. \label{fig:BarreteauJCG_Fig2}}
\end{figure}

\begin{figure}[t]
\includegraphics[width=\columnwidth]{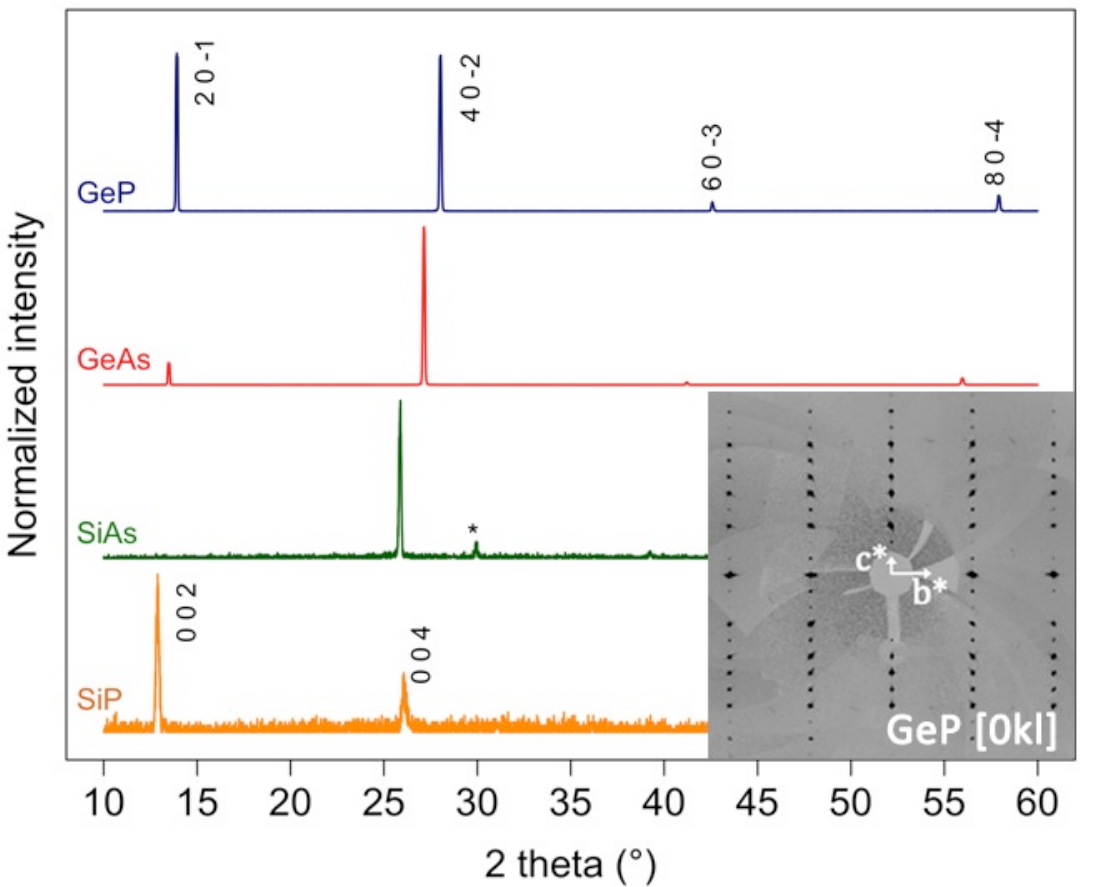}
\caption{XRD patterns: $\theta$-2$\theta$ diffraction from the
crystal surface of the four monopnictides (main panel), and
reciprocal space reconstruction in the [0\,\emph{k}\,\emph{l}]
plane from single crystal diffraction on GeP (bottom-right inset).
The secondary small peak labelled with a star in the pattern of
SiAs is due to the sample holder. The high background noise of the
SiP diffraction pattern is due to the very small size of the
sample. \label{fig:BarreteauJCG_Fig3}}
\end{figure}

Crystals cleave easily in the plane corresponding to the van der
Waals gap (see Fig.\,\ref{fig:BarreteauJCG_Fig2}) and clean powder
diffraction patterns, with a strong preferred orientation, were
obtained. This is confirmed by the $\theta$-2$\theta$ scans shown
in Fig.\,\ref{fig:BarreteauJCG_Fig3}, which are compatible with a
monoclinic symmetry \emph{C}2/\emph{m} with strong preferred
orientation along the [2 0 -1] direction. For GeP and GeAs, good
quality single crystals could be cleaved to confirm the crystal
structure (the reciprocal space reconstruction for the plane
[0\,\emph{k}\,\emph{l}] for GeP is shown as an inset in
Fig.\,\ref{fig:BarreteauJCG_Fig3}). The powder diffraction pattern
for SiP presents broader reflections but agrees with the
orthorhombic space group \emph{Cmc}2$_{1}$ with preferred
orientation along the [0 0 1] direction.

\begin{figure}[t]
\includegraphics[width=\columnwidth]{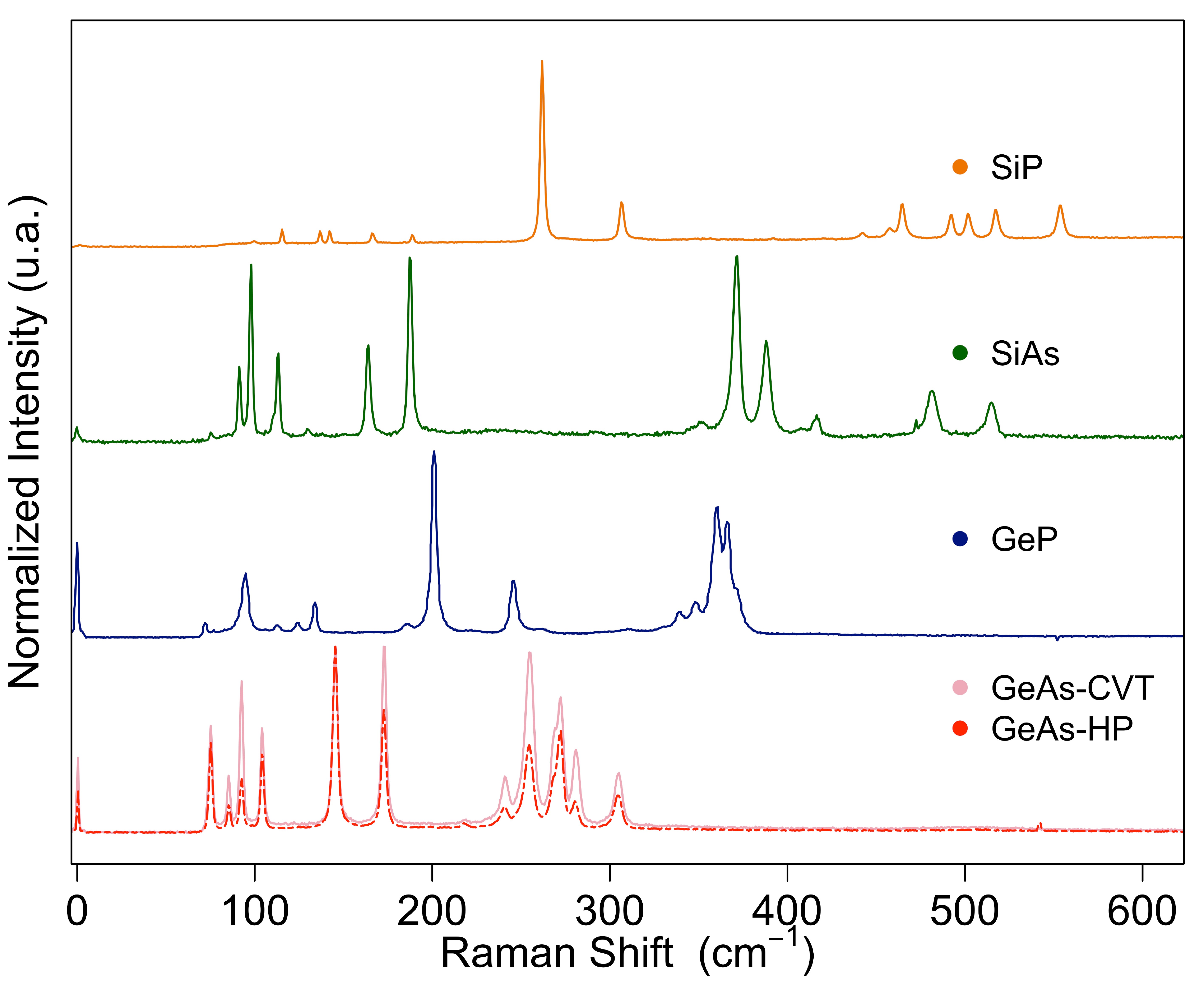}
\caption{Raman spectra of SiP, SiAs, GeP and GeAs (top to battom)
grown by HP. The bottom pattern compares GeAs grown by either HP
or CVT (red and magenta lines, respectively).
\label{fig:BarreteauJCG_Fig4}}
\end{figure}

The diffraction patterns obtained from the cleavage planes of the
two different structures in the same Bragg-Brentano geometry are
very similar. This is accounted for by the similar local symmetry
in the planes and the (Si,Ge)-(P,As) polyhedra that order in
similar chains in these planes, as described in ref. [11].

All samples were also characterized by Raman spectroscopy. The
narrow, well defined peaks of the Raman shift, as well as the very
low level of background proved the general good quality of the
crystals. Fig.\,\ref{fig:BarreteauJCG_Fig4} shows the Raman
spectra of SiP, SiAs, GeP and GeAs (from top to bottom,
respectively). The Raman shift in SiAs is in good agreement with
the previous Raman study reported by Kutzner et
al.\cite{Kutzner11}. The three compounds with a monoclinic
structure exhibit similar Raman spectra (similar groups of modes,
red-shifting when going from lighter to heavier compounds).
Unequivocally, the Raman shift of SiP is different from the
others, confirming the different crystal structure of SiP. The
bottom-most plot shows two patterns of GeAs crystals grown by
different techniques (HP and CVT): the reproducibility of the
Raman spectrum confirms the quality of the samples and the
reliability of the processing routes. Indexation of the Raman
modes of GeAs, GeP and SiP is not known at this stage. DFT
calculations of the phonon spectra are in progress. The Raman
study as a function of temperature shows no significant changes
down to 5 K, indicating that no structural transitions occur and
the symmetry is preserved.

\begin{figure}[t]
\includegraphics[width=\columnwidth]{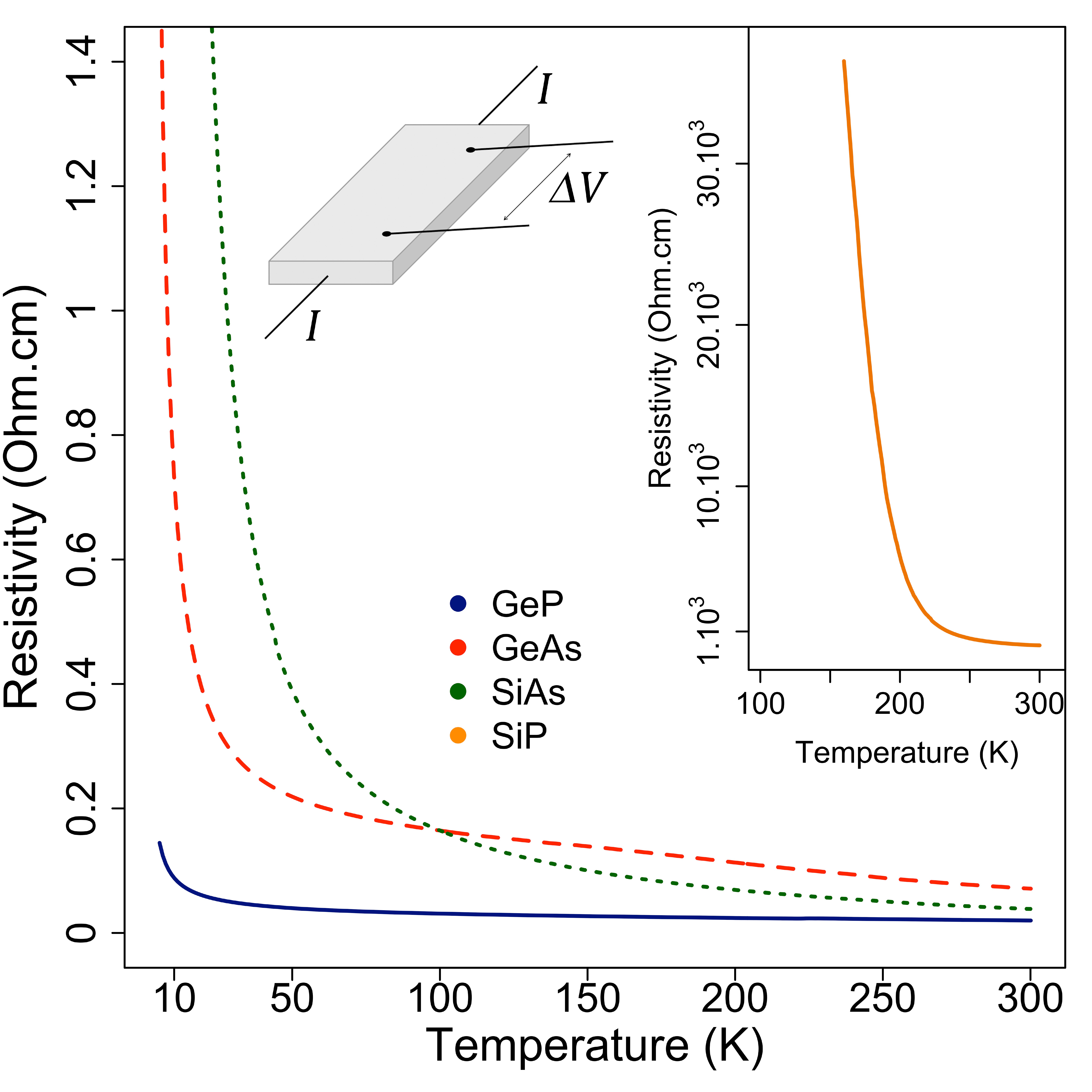}
\caption{Electrical resistivity as a function of temperature for
the four single crystals (darkblue: GeP, green: GeAs, red: SiAs
and, in the inset, orange: SiP). The drawing shows the standard
four-probe geometry used, with the current flowing in the cleavage
plane.\label{fig:BarreteauJCG_Fig5}}
\end{figure}

Electrical resistivity measurements were performed on several
HP-grown single crystals of each composition from room temperature
to 5K. Small deviations in the resistivity from one sample to
another with the same composition are consistent with the
difficulty in correctly estimating the thickness of these small,
layered crystals. As pointed out in
Fig.\,\ref{fig:BarreteauJCG_Fig5} , the four pnictides exhibit
semiconducting behavior. The values of electrical resistivity at
room temperature are in the same range for the monoclinic
compounds, GeP (0.02\,$\Omega\cdot$cm), SiAs
(0.038\,$\Omega\cdot$cm), GeAs\,(0.071 $\Omega\cdot$cm). On the
other hand, for the orthorhombic SiP the electrical resistivity is
four order of magnitude larger (141\,$\Omega\cdot$cm), as shown in
the inset of Fig.\,\ref{fig:BarreteauJCG_Fig5} ; the high
resistance of SiP prevents the complete characterization over the
whole temperature range. The large electrical resistivity of SiP,
as compared to the other members of the family, was reproducible
over samples from various batches and is likely to be related to
the structural difference between SiP and the other monopnictides.
No magnetoresistance effect is observed by repeating the $\rho$(T)
measurements under magnetic fields up to 7 tesla.

\begin{figure}[h!]
\includegraphics[width=\columnwidth]{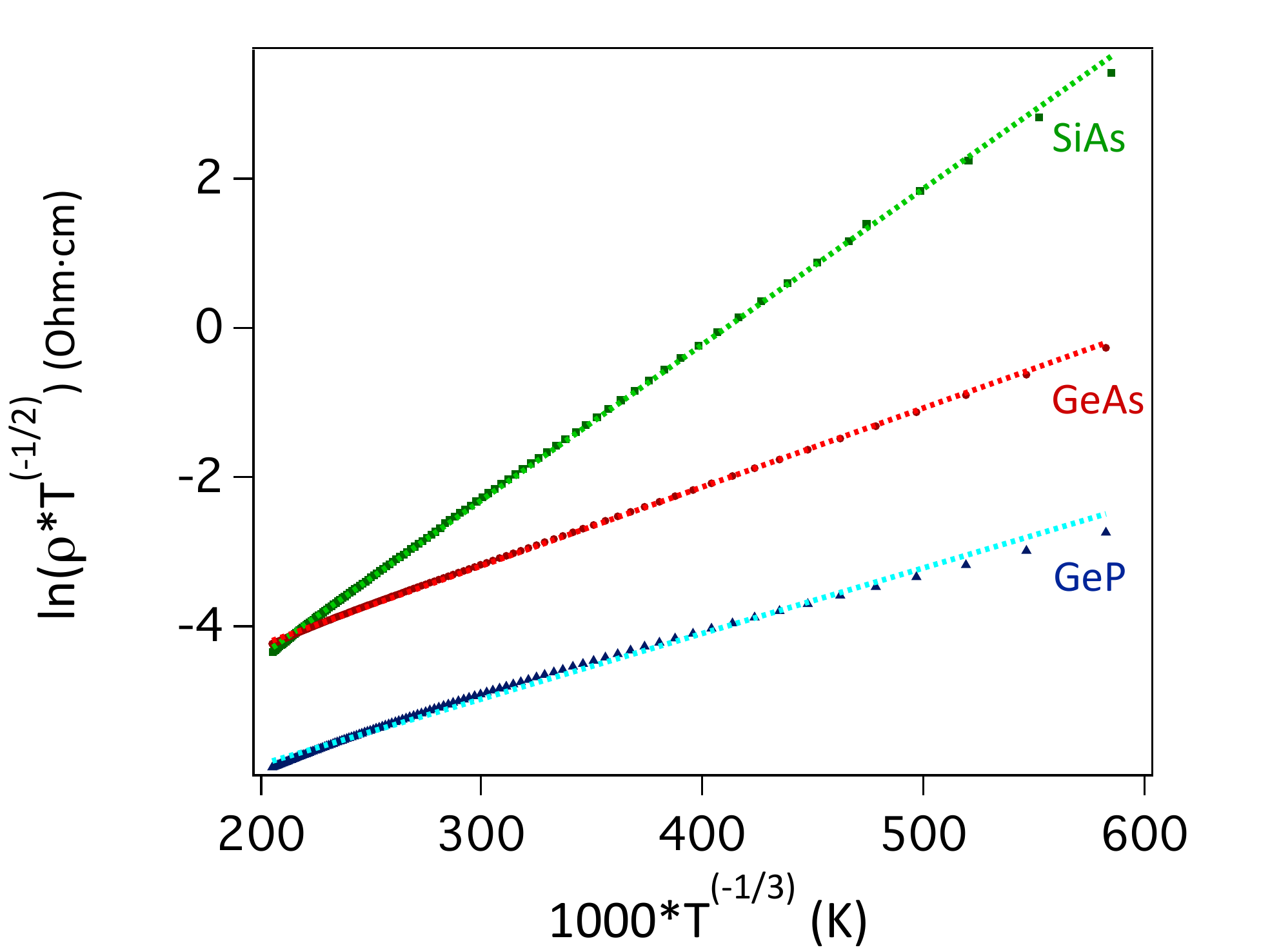}
\caption{Fit of the 2D Variable Range Hopping model to the
resistivity of GeP, GeAs and SiAs .\label{fig:BarreteauJCG_Fig6}}
\end{figure}

Despite their evident semiconducting behavior, a simple thermally
activated Arrhenius law does not fit to the experimental
resistivity of the monoclinic compounds. As a matter of fact the
dependence of the resistivity on temperature is very well
described by a Variable-Range-Hopping model \cite{Mott68}.
According to this model, the resistivity obeys the law
\begin{eqnarray}
\rho (T)= \rho {_{0}}\exp \left ( \frac{T_{0}}{T} \right
)^{1/(n+1)}
\end{eqnarray}

where T$_{0}$ is a constant and $\rho_{0}$ depends on temperature
as the square root of T \cite{Arginskaya94}. The factor \emph{n}
in the exponent indicates the dimensionality of the system, so
that the exponent 1/(n+1)=1/3 in a two-dimensional system. The
plot of ln($\rho$T$^{-1/2}$) versus T$^{(-1/3)}$ shown in Fig.5
for the three compounds GeP, GeAs and SiAs, evidences a linear
trend between T=5\,K and T=120\,K and agrees with a
Variable-Range-Hopping conduction in a 2-dimensional system. This
is consistent with the 2D-like structure of these materials and
confirms that the electrical conduction involves only the
(Ge,Si)-pnictide bonds in the quasi-2D slab oriented in the
[2\emph{l}\,0\,-\emph{l}] direction. This behavior is expected on
the base of the calculation of the Electron Localization Function
(ELF) in GeP reported by Lee et al. \cite{Lee15} that predicts the
absence of covalent bonds between the layers and strong covalent
Ge-Ge and Ge-P bonds in the layer.

The slope of the linear dependence of ln($\rho$T$^{-1/2}$) on
T$^{-1/3}$ shown in Fig.\,\ref{fig:BarreteauJCG_Fig6} is higher in
SiAs than in the two Ge-pnictides (for which it is the same). Such
slope, that is T$_{0}$ in equation (1), is proportional to the
hopping distance and inversely proportional to the density of
state at the Fermi level, \emph{N(E$_{F}$)} \cite{Arginskaya94}.
The higher T$_{0}$ would indicate that in SiAs the density of
growth-induced defects is lower. On the other hand, the equal
slope observed in GeP and GeAs VRH-linear regression suggests that
the origin of localized impurity states in the gap cannot be
ascribed to occupation vacancies or substitutional defects on the
pnictogen site, but are more likely related to the slightly
disordered local coordination of Ge, predicted by ELF calculations
\cite{Lee15}.

\section{Conclusions}
\label{4}

With the aim of searching for 2D layered semiconducting materials
that can be exfoliated down to atomically thin layers, we have
investigated the family of Si- and Ge-monopnictides (SiP, SiAs,
GeP, GeAs). Bulk crystals of these compounds were rarely grown,
had small size and never allowed systematic investigations of
their electronic properties. The crystal structure in which these
materials crystallized was object of controversy. In this work we
have shown that high pressure (in the GPa range) favors the
crystal growth of the four Si- and Ge-monopnictides. Those
containing Ge, in particular, can be grown with a large size (up
to 4-5\,mm$^{2}$ in the cleavage plane). Crystals of SiP and GeAs
could also be grown by the vapor transport technique, provided
that high-pressure pre-reacted elements were used as precursors
and I$_{2}$ was used as transport agent. We have confirmed the
monoclinic space group \emph{C}2/\emph{m} for SiAs, GeP and GeAs,
and the orthorhombic \emph{Cmc}2$_{1}$ for SiP. All compounds
exhibit a semiconducting behavior. Nevertheless, the electrical
resistivity of three of them (SiAs, GeP and GeAs) is found to
follow a 2D Variable Range Hopping conduction mechanism at low
temperature. These materials can be mechanically exfoliated and
the study of their properties as a function of the flake thickness
is in progress.

\section{Acknowledgements}
The authors gratefully thank J. Teyssier for his precious help in
Raman spectroscopy experiments. This work was partially supported
by the Swiss National Science Foundation through the "Sinergia"
project n. CRSII2-147607.



\end{document}